# Sinogram Space activity estimation for the IAEA PGET detector

**Author, company, and contact information**
Luca Presotto, PhD, is with the Nuclear Medicine Unit of IRCCS Ospedale San Raffaele, Via Olgettina, 60, 20132 Milano (MI), Italy. Phone number: +39 02 2643 5175. Email: presotto.luca@hsr.it
Ospedale San Raffaele is a research-focused hospital. The physics team of the Nuclear Medicine Unit has worked in the past in international collaborations for tomographic reconstruction (e.g.: PARAPET), it is a member of the EATRIS research infrastructure, and it has collaborations for the development of emission tomography reconstruction algorithms with major vendors.

## I. INTRODUCTION

THE PGET detector provides fine tomographic sampling of spent fuel assemblies (SFA). The data are provided in four broad energy windows and no detector normalization is provided. The two main tasks that the algorithm is required to fulfill is detecting missing pins and the quantification of the activity of the others. As the PGET detector uses the same principles as a SPECT camera, we believe that using algorithms for which robustness has been proven by decades of routine use in nuclear medicine is the optimal choice.

## II. THEORETICAL HURDLES

SFA are made of uranium oxide, which has a high cross section for photon interaction (mean free path: 1.5 cm @ 1.59 MeV, 0.7 cm @ 662 keV). Emissions from the pins at the center of the assemblies will be weak, and a very high background of photons that underwent Compton scattering will be present. As the presence of a pin can be detected only by measuring whether an excess signal is present over the background, the correct estimation of the Compton scatter is fundamental.

As the attenuation produced by a single pin is considerable, due to the small mean free path in $UO_2$, not knowing a priori which rods are present, makes the reconstruction challenging. Furthermore, the detector response factors ("normalization") are not measured. The simultaneous estimation of these 3 factors make the problem not convex, therefore as much prior knowledge as possible should be provided as input to the problem. Indeed, if the normalization factors are not spread randomly but they are correlated (e.g.: if all the pixels at the center of the radial coordinate have less sensitivity due to the high count rate), this effect cannot be separated from fluctuations due to the Compton scatter. To have full control on the corrections, we used only uncorrected raw sinograms.

## III. ALGORITHM DESCRIPTION

The algorithm is divided into steps that solve the problem with improving levels of approximation. In the first step, approximate scatter, normalization and attenuation are estimated. This estimate is used to determine the assembly geometry. With this available, an accurate estimation of activity and of all the corrections is performed in sinogram space. Finally, these corrections are used to perform an image reconstruction with a task optimized algorithm.

### A. Projectors and resolution modelling

A simple parallel projector was implemented for this algorithm. As the collimators are very long (100 mm), there is no need to model loss of resolution with depth as in conventional SPECT. Point spread function (PSF) was modeled using a shift invariant Gaussian convolution in sinogram space, tuned to the provided PSF measurements. Forward and back projectors were matched. For reconstructions with attenuation correction, the projectors accounted for depth-dependent attenuation.

### B. First approximate reconstruction

The normalization is estimated by comparing the measured sinogram with a version smoothed along the radial direction. The normalization is computed as the median of the ratio between the two along the angular direction. Bad pixels are recognized by measuring the difference of each line with its two nearest ones. These are not used in the subsequent reconstruction. To estimate the scatter, different energy windows are compared. It is imposed that the power in the frequency of the pin spacing in the sinogram is the same in the different energy windows, and the difference is retained as the Compton scatter. A first reconstruction is performed using a standard filtered backprojection (FBP) algorithm without attenuation correction. Chang attenuation correction is applied a-posteriori. With this information available, an approximate attenuation map is created by rescaling the emission image. A higher quality image reconstruction is then performed using a statistical method, which also guarantees cold contrast recovery in very few iterations[1], leading to the image which is then used for the geometry detection.

### C. Object geometry recognition

It is known a priori that the geometry can be only either square or hexagonal. This is recognized by performing a Fourier transform of the sinogram along the radial direction and counting the number of peaks along the angular direction. From the initial qualitative reconstruction, the centers of all the non-empty rods are detected. From the position of all the rods and their spacing a regular grid is created. At this stage, the average diameter of the rods is also estimated. This is achieved by measuring the profiles in sinogram space along the angular directions parallel to the object orientation, accounting for the detector PSF.

### D. Sinogram space quantification

To achieve the most accurate quantification possible, we decided to work directly in sinogram space. The gradient of the likelihood of the tomographic problem, when the exact Poisson model is used, is

$$\frac{\partial L}{\partial \lambda_j} = \sum_i c_{j,i} \frac{y_i - \left(\sum_\xi c_{i,\xi} \lambda_\xi + s_i\right)}{\sum_\xi c_{i,\xi} \lambda_\xi + s_i}$$

where, $c_{i,j}$ are the projection matrix factors, accounting also for detector normalization, attenuation, and PSF, $\lambda_\xi$ is the $\xi$-th pixel to be estimated, $s_i$ is the Compton scatter estimate for sinogram bin $i$, and $y_i$ is the measured sinogram element for bin $i$. In the problem under analysis, however, we are interested only in estimating the activity in a limited number of regions of interest (i.e.: one for each pin). In this case, we can explicitly compute $c_{i,j}$ for each pin, by forward projecting its binary mask. Then, the previous activity update equation becomes computationally inexpensive and can be iterated ad libitum, as no further forward and back projections are needed. Furthermore, as the system matrix in this case can be written out explicitly, the exact statistical error on the activity quantification can be computed by inverting the Hessian matrix of the problem, which is only of size $N_{PINS} \times N_{PINS}$.

### 1) Energy window used and attenuation correction

The two most active isotopes in SFA are $^{154}$Eu and $^{137}$Cs. Cesium is by far the most active but its energy is only 662 keV where the mean free path in $UO_2$ is about 0.7 cm. In a $10 \times 10$ assembly this means that the pin at the center will be on average attenuated by a factor $\approx 200$. This makes the signal from the inner pins in SFA negligible compared to the uncertainties in the calibration factors and in the Compton background, an issue that cannot be solved by increasing the count statistics. We therefore worked in the energy windows containing only the highest energy photons of $^{154}$Eu. Specifically, 1596, keV, with $\approx 1.7 cm$ mean free path, but only 1.79% abundance, the 1274 keV peak, with a 35% abundance and $\approx 1.4 cm$ mean free path, and the two peaks around 1MeV with a combined 28% abundance but only $\approx 1.1 cm$ mean free path. As we are limited to very large energy windows, we chose to work with either the highest energy window, if it contains enough counts for a high quality reconstruction, or with the second highest. When the highest window is 1 MeV- 3 MeV, we use only this one. When the highest energy window is 1.5 MeV – 3 MeV we use it exclusively if sufficient counts have been collected, allowing to achieve less than 3% statistical error on the activity of the central rod. If this is not possible, the second highest energy window is used. For mock $^{60}$Co reconstructions, the energy window containing the two photopeaks is used (0.7 – 1.5 MeV).

As different gamma peaks experience different amounts of attenuation, we model this phenomenon exactly by computing one forward projected mask in sinogram space for each gamma energy, and they are weighted both by the abundance and by the detection probability (both in the photopeak and in the Compton tail). Appropriate attenuation coefficients are used for each gamma energy, using tabular values from the NIST database, both in $UO_2$ and in water. If a single attenuating coefficient is used, the contribution of the inner rods to the sinogram will be over- or underestimated and, given the very large amount of attenuation present in this setup, their activity will not be measured correctly.

### 2) Scatter and normalization factors update

The activity reconstruction loop is nested with another loop that updates that scatter and the normalization corrections. The scatter estimate is updated as the difference between the current activity estimate and the measured one, which is then convolved by a heavy filter that retains only the frequencies and the directions allowed by the physics of the detection and scattering process. The normalization factors are updated using the median, along the angular direction, of the ratio between the measured and the estimated sinogram. As the whole problem is not convex, these updates are performed in a relaxed fashion, applying only a fraction of the update at every iteration.

### 3) Empty pins detection

After every calculation of the activity, empty rods are detected as those having less than 66% activity of the median of the distribution or those which activity value is an outlier compared to the distribution of the other pins (i.e.: more than 3 standard deviations away from the mean).
When an empty pin is detected the sinogram masks for each pin are computed again, accounting for this updated information concerning the attenuation. The previous thresholds can be adjusted to achieve the desired balance between false positive and false negative detection rates.

### E. Final image reconstruction

The previous step, on top of providing the activity estimation and its error for every pin, provides updated scatter and normalization corrections. This information is used to provide a final high quality reconstruction, so that the user can qualitatively assess whether the previous procedure provided reasonable results, using 1 mm pixels. The reconstruction is performed using a preconditioned gradient descent algorithm modelling the full Poisson statistics of the detection process. A special preconditioner that guarantees almost immediate convergence is used [1]. This is favored over the more common MLEM to fully recover cold contrast between the rods with very few projection operations. All corrections (normalization, attenuation, Compton scattering) are included in the statistical model, and not precomputed, to guarantee minimal noise [2]. To improve image quality, a denoising prior that guarantees unbiased uniform resolution smoothing is added to the likelihood function [3].

### F. Rods classification task

Mock $^{60}$Co acquisitions feature pins that belong to 5 different activity levels. However, these groups partially overlap and their distribution spans only a $\pm 10$ range. Accounting for the systematic uncertainties in the estimation of the activity of each rod, which in the training phase we estimated to be around 10%, this makes it impossible to use data-driven clustering approaches. Therefore, we only scale the reconstructed activity distribution so that it overlaps with the known one of the 5 groups. Then, each rod is associated to the closest cluster, knowing a priori its median value.

## IV. COMPUTATIONAL COMPLEXITY

The proposed algorithm does not require special computing resources. The image reconstruction part uses optimal techniques that require very few projection operations for full convergence. The only expensive computational part is the generation of the masks in sinogram space for the sinogram reconstruction. Furthermore, they need to be recomputed each time that an additional empty pin is found. The computation time therefore depends on the number of pins, on the number of energy peaks modeled, and on the number of times these masks need to be updated. The current code, which is prototypal, not parallelized, and implemented on a CPU, requires for the full reconstruction between 7 minutes (e.g.: competition dataset 4, 100 pins) and 25 minutes (e.g.: competition dataset 6, 288 pins) on a 2017 desktop computer with an Intel i7-7700K CPU. A dedicated parallelized implementation on a GPU should reduce the computation time to less than 1 minute. A modification of projectors, to allow the update of the attenuation maps without having to reproject the masks in sinogram space, should further reduce the computation time notably.

It should be noted that performing the activity quantification in sinogram space allows running the equivalent of many thousands of conventional iterations with minimal computational requirements.

## V. ANALYSIS OF THE LIMITATIONS, EXPLICIT AND IMPLICIT ASSUMPTIONS, AND POSSIBLE IMPROVEMENTS

### A. Pin positions and size

The sinogram space reconstruction can be performed only if the exact position and exact diameter of each pin is known. A single fixed pin diameter is used for all the pins in a reconstruction, and it is derived from a fit in sinogram space. As we are working with 2 mm pixels, which is the resolution of the detector, variations of $\pm 1 mm$ of the diameters between individual pins are negligible, and variations of $\pm 2 mm$ should have tolerable impact.

From the information provided before the challenge, we generate the map of the pins assuming them to be uniformly spaced on a regular grid. For the same reasons, we expect irregularities of $\pm 1 mm$ to be negligible, and of $\pm 2 mm$ to be acceptable. With all the setups provided in the training and competition phase, we never had the reason to suspect misalignments or variations in rod diameters larger than $\pm 1 mm$.

### B. Scatter and normalization correction

Due to the presence of heavily scattering material, this is the most crucial step of the algorithm. After the initial estimation, we update it using the difference between the current reconstruction and the measured sinogram, filtering this difference to retain only the directions and the frequencies physically allowed by the Compton scattering process. The ratio, instead, is used to update the normalization factors. We initialize this iterative process under the assumption that all the pins have equal activity. This process is restarted whenever an empty pin is found. This assumption could be disputable in SFA, where in principle pins in some regions might have activity much higher than the others. However, this is the safest assumption given the current knowledge, and it is less strict than the one used by IAEA in the PGET detector report to estimate the normalization factors.

### C. Empty pin detection

The detection of empty pins is currently performed assuming that all the pins have similar levels of activity. This can lead to potentially false detections/missed detections in case that the true activity distribution is very broad. However, unless additional prior knowledge is given, this assumption is the safest. The criteria of activity difference with the nearest pins was not found to perform better, as it might fail when multiple empty pins are present in a small region.

When an empty pin is found, it is removed from the attenuation map. This results in very low levels of activity estimated in the pins recognized as empty, which are potentially artefactual if a true pin was mistakenly considered missing. A different approach might be estimating the final activity, to be reported in the output table, assuming that the attenuation is present even where the pins are known to be absent. However, this can lead to artefacts in the reconstruction when multiple pins are missing, especially if they are close together. The choice between these two options is left to the user, according to the preferred kind of bias.

### D. Discarded options and possible improvements

The author has a large experience in joint estimation of activity and attenuation in emission tomography [4]. In a recent work, an algorithm to increase robustness of this problem was introduced [5-6]. If both data could be estimated from emission data alone, independent confirmation of missing pins could be provided. However, these algorithms are badly conditioned, and they have been shown to diverge when even minimal errors are present in the scatter or normalization correction [7], both of which are unknown here. Acquiring multiple narrow energy windows for each photopeak, to exploit the high energy resolution of the CZT detector, would allow a robust estimate of the Compton scatter background. This is particularly important to correct the low-angle scatter, which follows the signal more closely, as the large-angle one can be easily modeled and, in any case, it does not bias reconstructions locally. Furthermore, as the signal in the "scatter windows" is low frequency, it would provide an independent calibration for the normalization. As can be seen by statistical errors obtained reconstructing the signal in a single energy window (~1% in the center of the assembly), the counting statistics even in narrow energy windows would be sufficient to estimate activity with high degree of confidence, without having to resort to the many corrections that had to be introduced here. Furthermore, the current scheme used to account for different energy peaks in a single window, can be used as is for joint reconstruction with multiple windows, one for each peak. In such case, estimation would be orders of magnitude more robust and it might even be possible to estimate the $^{154}$Eu to $^{137}$Cs ratio for each pin.